# The relationship between junctional zone, smooth muscle tone and intra-uterine pressure

Nicolas R. Chevalier[1], Emilie Huon[1], Julie Poujol[2], Louise Bandiera Mezei[2], Henri Souchay[2], Edouard Poncelet[3]

[1]Laboratoire Matières et Systèmes Complexes, Université Paris Cité, CNRS, F-75013 Paris, France

[2]GE Healthcare, Buc, 78530, France

[3]Service de radiologie, Centre Hospitalier Valenciennes, France

***Correspondence:**
Nicolas R. Chevalier
nicolas.chevalier@u-pariscite.fr

## Abstract

The junctional zone is a thin low contrast region of the inner myometrium, visible on T2-weighted MRI images. Investigators have long sought histologic particularities of this region. Here, we summarize evidence that this region's specific appearance is not due to a particular tissue composition, but is a reflection of underlying smooth muscle activity. Smooth muscle tone in the spirally fibered myometrium wrings out blood and interstitial fluids to create a static low-contrast T2 region around the myometrium, while phasic smooth muscle contractions create transient, propagative low-contrast waves. We show that the competition between centripetal inhibitory adrenergic innervation from the serosa and centrifugal excitatory hormonal gradients from the endometrium divides the uterus in a tonic inner- and a quiescent outer-region. Bridging MRI and manometry studies, we discuss the relation between smooth muscle activity and intra-uterine pressure. Dysregulation of nerve and/or hormonal activity can result in a thicker tonic smooth muscle layer, entailing a thicker junctional zone and high intra-uterine pressures characteristic of dyscontractility in severe dysmenorrhea, endometriosis and adenomyosis.

# Main text

### Introduction

Dyscontractility is the cause of pain in dysmenorrhea (Hellman et al., 2018; Woodbury et al., 1947), and mounting evidence indicates that it represents a primary etiological factor in the development of endometriosis (Sampson, 1927; Vercellini et al., 2024) and adenomyosis (Chevalier et al., 2026; Leyendecker et al., 2015) . Early investigators revealed uterine contractions in the non-pregnant uterus by manometry (Moir, 1934; Smith, 1984) . A very informative but today to some extent forgotten literature has examined intra-uterine pressure abnormalities in dysmenorrhea and their relation to pain (Filler & Hall, 1970; Woodbury et al., 1947). Following the development of magnetic resonance imaging (MRI) and transvaginal ultrasonography (TVUS), it became possible to study uterine contractility spatio-temporal characteristics across the menstrual cycle (Lyons et al., 1991; C. O. Rees et al., 2023) and in pathological conditions (de Boer et al., 2021). The factors influencing contractility are numerous: patient age, menstrual cycle phase, pathological condition, medical treatment. Depending on the method used – manometry, static-, cine-MRI or -TVUS – investigators have reported and focused on different observables: basal pressure, peak pressure, contraction frequency, direction, junctional zone (JZ) thickness or dynamics. As a result, the literature available today is rich, but fragmented. Getting an overview of the dependance of any of these observables on, say, menstrual cycle phase, requires collecting and stitching data from different studies performed on different patients. In addition, the relationship between what is measured by MRI and by manometry is rarely, if at all discussed.

Our aim here is to perform a synthesis of the existing literature, but also to bring forth novel arguments that allow us to draw a consistent scheme of the nature of the JZ, of the physiological reason why only this thin band of myometrium is tonic, and of how this relates to intra-uterine pressure. We start by describing data on static and dynamic characteristics of the JZ, on time scales encompassing a woman's life (years), the menstrual cycle (month), and individual contractions (seconds). We then explain why the JZ should not be considered an anatomical structure, but as a reflection of smooth muscle activity (tone). This argument had already been formulated 40 years ago (Lee et al., 1985) , we add here original elements to support it. We then present, on the basis of our own recent findings (Chevalier et al., 2026; Huon, Zig, et al., 2026), how the competition of outside-in radiating adrenergic nerve inhibition and inside-out diffusing hormonal smooth muscle activation can give rise to a steep, well demarcated JZ. We finally discuss the relation between smooth muscle tone, the JZ and pressure as measured by intra-uterine manometry.

### MRI characteristics of the JZ

The JZ is a low contrast region of the inner myometrium, adjacent to the endometrium, visible on T2-weighted MRI images of fertile women. The JZ is intrinsically hormone dependent: it is indistinct in pre-menarche girls and post-menopausal women (Fig.1A) (Novellas et al., 2012);

oral contraceptives cause the JZ to drastically thin down (Kido et al., 2005) . The JZ is slightly thicker in women in their 30s than in women in their 20s (He et al., 2016) . The JZ changes with the menstrual cycle (Fig.1B). The most prominent changes in JZ morphology occur on the first day of menstruation (Kataoka et al., 2005) . Studying a cohort of healthy and dysmenorrhea patients, these investigators found that the JZ on the $1^{st}$ day of menses often encompassed half of myometrium thickness (i.e ~12 mm), sometimes even the entire myometrium. This thickening was more pronounced in dysmenorrhea patients. The dysmenorrhea group also presented more often with a bulging JZ, i.e. a non-uniform thickness. JZ thickness and non-uniformity were both found to decrease sharply on day 2 of menstruations, and reached the commonly cited thickness of 4-8 mm at day 3. JZ thickness was found to increase during the remainder of the menstrual cycle in two studies: from 4 to 7 mm (Wiczyk et al., 1988) and from 4.5 to 5.5 mm (He et al., 2016). One study did not find a significant variation (Meylaerts et al., 2017), but only examined the mid-follicular, ovulatory and mid-luteal period. We think that all of these results can be reconciled by considering that thickness variations are most pronounced on day 1 of menstruations and may go undetected if measurements are performed on later days.

On shorter time scales, phasic contractions of the myometrium (Hellman et al., 2018; Masui et al., 2001; Togashi, 2007; Togashi et al., 1993) also give rise to dynamic changes in JZ morphology (Fig.1C). These have been classified (Togashi, 2007) as either “bulging” contractions usually involving the entire myometrium, or more subtle peristaltic wave-like motion confined to the endometrium and its vicinity. Both movements type result in a transient decrease in T2 (darkening), that propagates either in the cervico-fundal or fundo-cervical direction (Masui et al., 2001; Nakai et al., 2003) . This transient darkening also radiates in deeper layers of the myometrium. Fig.1 and Video S1 show an example of this wave-like, dynamic T2 change. Thus, and as previous investigators have warned, clinicians have to make sure that transients that “mascarade as leiomyomas” (Kido & Togashi, 2016) do not bias their morphological assessment of JZ morphology.

A thickened JZ is found in several highly prevalent uterine conditions. As already mentioned above, it was found to be thicker during menstruations in dysmenorrhea patients (Kataoka et al., 2005). A JZ thicker than 10-12 mm (measured outside of the menses period) is a strong indicator of adenomyosis (Novellas et al., 2012). The posterior wall JZ is also thicker in endometriosis patients (Kunz et al., 2005) and in patients with severe dysmenorrhea following endometriosis lesion surgery (Parker et al., 2006).

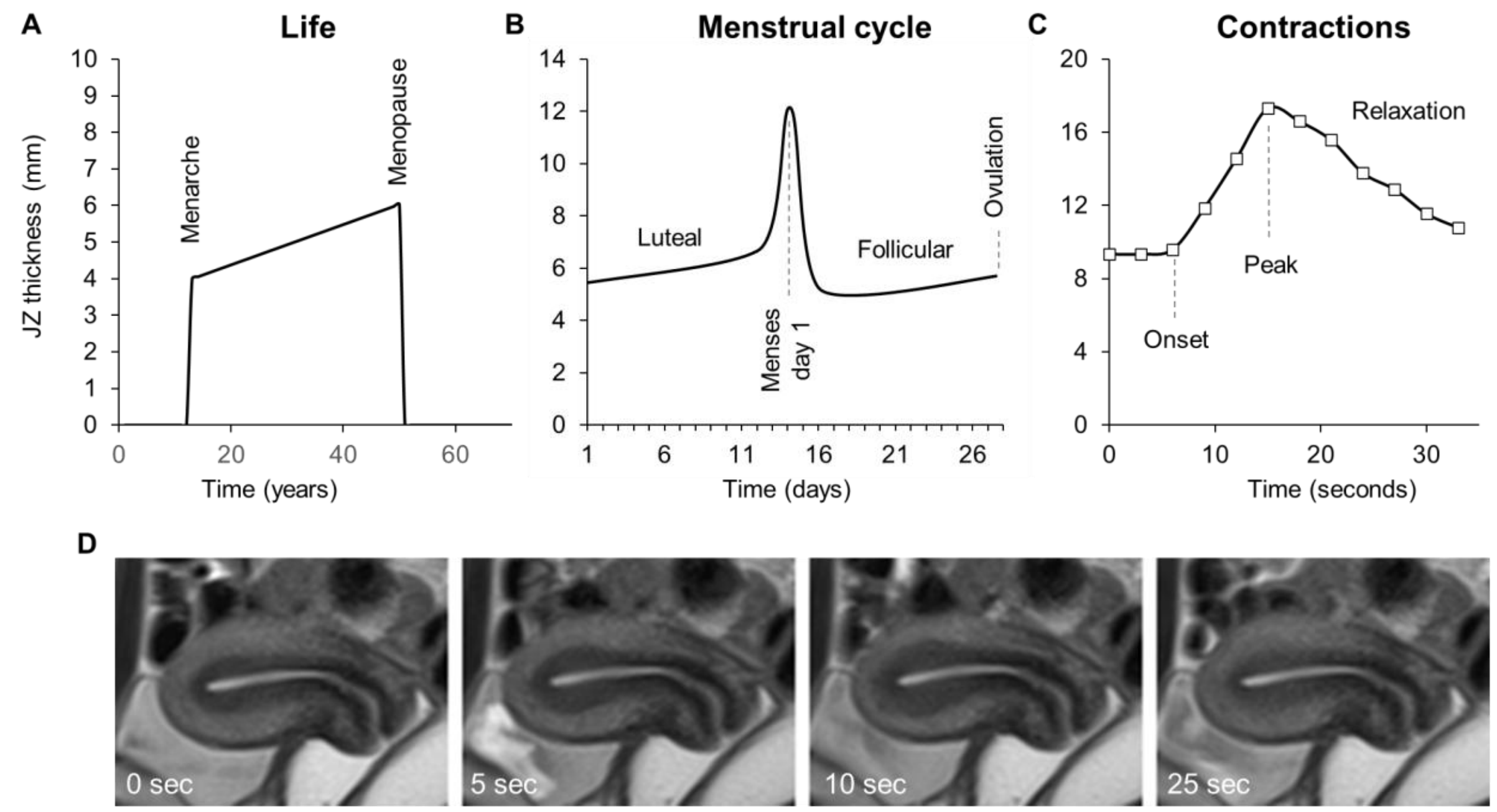

*Figure 1. JZ thickness variations during a woman's life, her menstrual cycle, and at each myometrium contraction. The variations illustrated in A and B are schematic and the reader is referred to the relevant literature for details. (A) The thickness varies from menarche to menopause from ~4mm to ~6 mm (He et al., 2016; Kiguchi et al., 2017). IRM performed on 9 pre-menarche girls did not reveal a distinct, measurable JZ (Demas et al., 1986) . A JZ was detected in only 4/28 postmenopausal patients (Kiguchi et al., 2017). In both pre-menarche and post-menopause, the inner myometrium has lower T2 intensity than the outer myometrium, but signal variations are often too small and restricted to a thin region to allow for quantification. (B) JZ thickness increases at day 1 of menstruations: it can encompass half of myometrium thickness (~12 mm) in healthy women with mild or no pain on menstrual cycle 1, and up to the full myometrium thickness in women with moderate-severe pain (Kataoka et al., 2005) . The thickness decreases to its basal level by day 3 of menses (Kataoka et al., 2005). Evolution of JZ thickness in follicular and luteal phases is depicted as an average from two studies (He et al., 2016; Wiczyk et al., 1988). Although no data is available at pre-menses days, the increase in JZ thickness likely occurs within a brief period of time. (C) Increase in JZ thickness during a uterine contraction, measured from Video S1. Similar findings have been previously reported (Hellman et al., 2018; Masui et al., 2001). (D) Cine MRI showing the enlargement of the JZ during a contraction (0, 5, 10 sec) and the uterus relaxing back to its initial state (25 sec).*

## The JZ is a physiologic phenomenon: fluid squeezing by myometrial smooth muscle

The first researchers who observed the JZ on static MRI images thought that it would reflect a particular anatomic structure of the myometrium. The plethora of histological studies (Habiba et al., 2026; Harmsen et al., 2023; Lee et al., 1985) that strived to identify distinct features of this inner myometrial layer have however all reached the same conclusion: there is no "banding" on histological sections. All anatomical changes from outer to inner myometrium are graded, they develop smoothly over the whole thickness of the organ, with no abrupt change in cell nature, density or fiber orientation that corresponds to the JZ (Chevalier et al., 2026; Harmsen et al., 2023; Scoutt et al., 1991).

Following these observations, Lee et al. first mentioned the possibility that the JZ may result from a physiologic phenomenon rather than a morphological one (Lee et al., 1985) . Their conclusions were based on the fact that ex-vivo, post-hysterectomy uteri presented with much weakened and sometimes absent JZ features. Similar observations were made in another study (McCarthy et al., 1989). This suggested that the JZ could be related to organ *activity* (e.g. blood flow) rather than structure.

This hypothesis was bolstered with the advent of cine-IRM and the first description of JZ dynamics (Fig1C,D & Video S1): a low-intensity region that fluctuates within seconds after each peristaltic contraction and within days at each menstrual cycle could obviously not be due to morphology. T2 is determined by the mobility of water within tissues: while hydration shell low-mobility water has T2 on the order of 5 ms, vascular blood is in the range 200-300 ms and pure water ~4000 ms (Sinha & Sinha, 2024). A low T2 value in the JZ indicates a loss of overall tissue hydration. Togashi and co-workers were the first to emit a simple, but in our opinion not sufficiently known and accepted hypothesis (Togashi, 2007): blood being squeezed out of the myometrium by smooth muscle contractions could cause a locally lower T2. This explanation needs to be expanded upon.

First, it is consistent with basic smooth muscle biology. Active smooth muscle presents a static force component (tone), and a phasic one (rhythmic contractions) (Fig.2A). Similarly, the JZ presents a static component (its thickness as measured by the clinician), and a dynamic one (rhythmic propagating T2 changes visible on cine MRI, Fig.1C, Video S1).

Second, both MRI (Kunz et al., 1996) and TVUS (Birnholz, 1984; de Vries et al., 1990; Lyons et al., 1991) have consistently shown that only the inner myometrium is strained (deformed) during rhythmic uterine contractions. This is consistent with the idea that smooth muscle is active (tonic) in the inner myometrium within or close to the JZ, but not in the outer myometrium.

Third, not only blood, but also extracellular fluid can be expelled as a result of smooth muscle tone and contractions. T2 values in the JZ and outer myometrium are respectively ~50 ms and ~60 ms (He et al., 2016; McCarthy et al., 1989). Based on a study on the relation between T2 and hydration in the intervertebral disc (we are not aware of similar studies in muscle) (Marinelli et al., 2009), we find that this corresponds to an approximately ~2-3% lower water content in the JZ compared to the outer myometrium. Using reported T1 values (McCarthy et al., 1989) and the Fatouros Marmarou model for brain tissue (Fatouros et al., 1991) yields a similar difference. A 1.7% hydration difference between the JZ versus the outer myometrium has been measured on ex-vivo uteri (McCarthy et al., 1989): it probably underestimates the in-vivo value, since muscle tone and blood content are both affected by organ ablation.

Fourth, the squeezing effect may be particularly prominent in the uterus because of the particular weaving of smooth muscle fibers in this organ: histology (Goerttler, 1968) and diffusion tensor 3D MRI (Weiss et al., 2006) have shown that most muscle fibers in the myometrium are arranged in counter-winding spirals (Fig.2B). When such fibers contract, they expel water, just as in the everyday experience of wringing out a piece of wet cloth (Fig.2C).

Fifth, one may think that this wringing/squeezing effect could have been reported in other smooth muscle organs upon cine-MRI imaging. One should however here remember that the myometrium is by far the thickest smooth muscle in the human body, with a wall width of ~25 mm. In comparison, the second thickest smooth muscle, the tunica muscularis of the colon, measures only ~2 mm. With current cine-MRI voxel resolution, the colonic tunica would appear as roughly ~2-5 pixel thick, making it difficult to distinguish any water dynamics during contractions in this, or other smooth muscle organs.

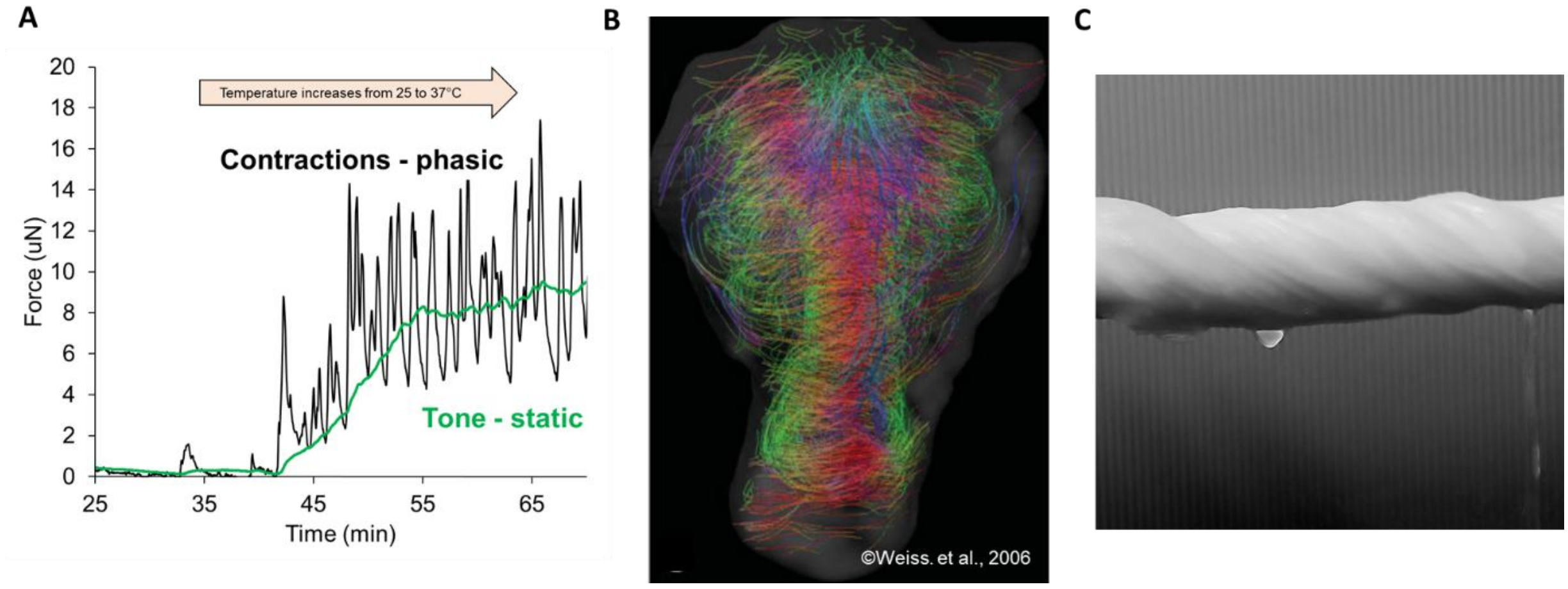

***Figure 2.*** *A) The force developed by smooth muscle consists of a quasi-static component, tone (green), and a phasic component, the rhythmic contractions (black). Similarly, the JZ features a static, and a dynamic component. This particular curve was obtained by measuring the force produced by a ring of fetal chicken intestine during thermalization from 25°C to 37°C. B) Diffusion-tensor MRI image reproduced from Weiss et al., 2026: uterine smooth muscle fibers are arranged in spirals around the lumen, C) When wringing out a piece of wet cloth, it forms spirals and flushes out water, similarly to the myometrium when it contracts.*

For all these reasons, we believe that the JZ should by now be accepted as resulting from blood and extracellular fluid depletion by the tonic and phasic contractile activity of the inner myometrium. All these considerations do not however explain why smooth muscle is only tonic in a well demarcated region of the inner myometrium, and not in the outer myometrium: we address this below.

## The JZ as a result of the competition between centripetal nerve inhibition and centrifugal hormonal excitation

We recently reported (Huon, Zig, et al., 2026) that, in an organ bath, the entire muscle thickness spontaneously contracts, not just the inner myometrium, or the JZ for that matter. In this experiment, nerve connections have been severed and nerve input is lost: the contractions are insensitive to the nerve blocker tetrodotoxin (Huon, Rayssac, et al., 2026), they are purely myogenic. This result and shows that the entire myometrium presents a uniform, spontaneous *myogenic potential*.

In another recent report (Chevalier et al., 2026) we have shown that sympathetic adrenergic nerve fibers present a centripetal organization: they are most dense at the serosa, and gradually become sparser as they radiate inside the myometrium. These tyrosine-hydroxylase-positive fibers are almost absent in the myometrium immediately adjacent to the endometrium. These nerve fibers produce adrenaline, which binds to β2 adrenergic receptors on SM, inducing its relaxation via an intracellular cAMP increase. This mechanism is common to vascular and bronchial SM, and is the reason why salbutamol agonism of β2 adrenergic receptors is an effective relaxant of all these muscle types (Lalos & Joelsson, 1981). Importantly, we found that adrenergic nerve fibers are significantly depleted in the inner myometrium of adenomyosis patients compared to controls (Chevalier et al., 2026). This finding obviously resonates with the thickened JZ observed in adenomyosis: the tonic smooth muscle layer is thicker in this disease.

These two ingredients – a uniform myogenic potential and a centripetal muscle inhibition - already provide a plausible explanation of why SM tone is restricted to the inner myometrium layers. It does not however explain JZ thickening at menses (Fig1.B). Although adrenergic innervation has been shown in rats to present some plasticity throughout the menstrual cycle (Zoubina et al., 1998), it is unlikely that this plasticity can account for the relatively quick changes in JZ thickness (Fig.1B). A third, hormonal ingredient is necessary to explain this. The prostaglandin PGF2α and its metabolite 15-keto, 13,14-dihydro PGF2α very likely play an important role: these pro-contractile hormones are known (Lundström & Green, 1978) to be enriched by a factor $10^3$-$10^4$ in the endometrium and menstrual fluid compared to serum levels, and to be present at higher levels in dysmenorrhea patients. By activating contractions, PGF2α allows the evacuation of menses (Ruan et al., 2011): this means that the motility of the uterus depends on the chemical nature of its content, much like what happens in the intestine (Fung et al., 2025). PGF2α can directly diffuse in the myometrium from the endometrium and uterine cavity: since typical diffusion times are ~10 mm in 24h, prostaglandins could reach distances comparable to JZ thickness in the time window corresponding to the onset of menses. Following this initial surge, the rapid decrease in PGF2α concentration in menstrual blood in the following days (M. Rees et al., 1984) would have the JZ return to its normal thickness.

We suggest that the profile of SM activity across the myometrium results from the combined effect of these three activities: a uniform myogenic potential, an outside-in (serosa-to-endometrium) nerve inhibition, an inside-out (endometrium-to-serosa) hormone excitation. Smooth muscle tone is a non-linear function of local adrenaline (inhibitor) and PGF2α (activator) levels (Goutelle et al., 2008). Fig.3 shows how the respective spatial distributions of these molecules can give rise to a step-like tone profile similar to that of the JZ.

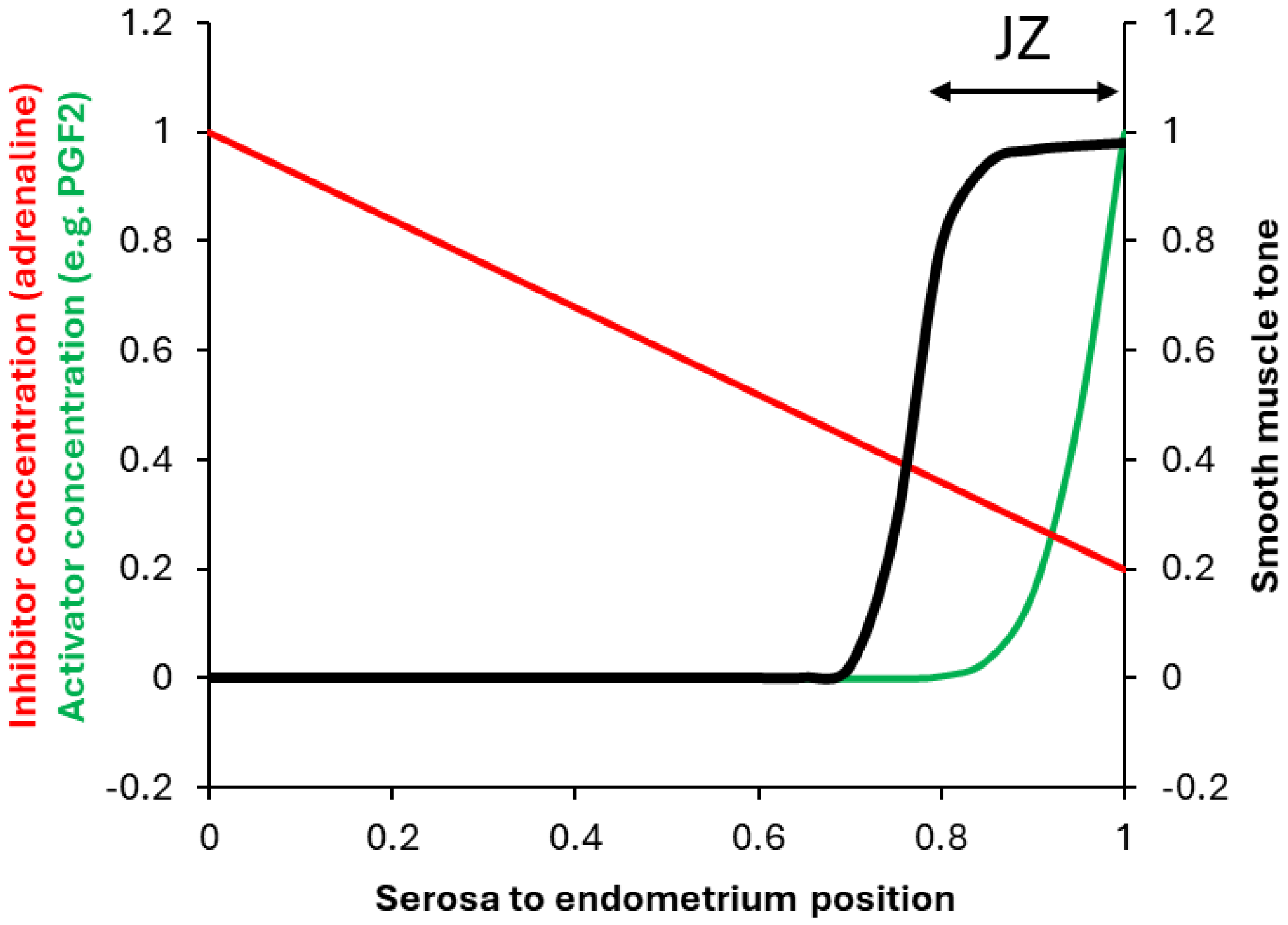

***Figure 3. A step-like smooth muscle tone profile can emerge from the centrifugal distribution of nerve fibers in the myometrium and the centripetal diffusion of excitatory hormones from the endometrium****. Concentrations and tone here have arbitrary units. The adrenaline profile in red reflects the distribution of adrenergic fibers in the myometrium (Chevalier et al., 2026) . The PGF2α profile in green is a diffusion-type complementary error function. The smooth muscle tone T in black was computed from these two profiles using a non-competitive activator-inhibitor Hill equation* $T = \left(\frac{A}{K_A+A}\right)\left(\frac{K_I}{K_I+I}\right)$*, with A and I the activator and inhibitor concentrations, and* $K_A$ *and* $K_I$ *the activator and inhibitor dissociation constants; we chose* $K_A = 10^{-3}$ *and* $K_I=10$ *here.*

## The uterus as a pressure vessel

The link between the JZ and smooth muscle tone naturally invites us to see how JZ features are related to intra-uterine pressure. Intra-uterine catheters (or balloons) have historically been the first modality by which uterine contractions were measured, in 1872 in the pregnant- and in 1925 in the non-pregnant uterus  (Smith, 1984). Early research has soon thereafter found that dysmenorrhea was associated to heightened peak intra-uterine pressures. Pain was felt only during pressure increases, although not all pressure increases resulted in pain (Woodbury et al., 1947). The pain-contraction association has been revisited by MRI (Hellman et al., 2018). The last to our knowledge published manometry study has found that endometriosis patients present 2-3 times higher basal & peak pressure levels and contraction frequency compared to healthy women (Bulletti et al., 2002). We are not aware of any manometry studies pertaining to adenomyosis.

During the menstrual cycles of healthy women, pressure waves are infrequent (~1 cpm) but of high amplitude (~100 mmHg) at menses, and frequent (~3 cpm) but of low amplitude (~10-20 mmHg) at ovulation (Bulletti et al., 2000; Eskes et al., 1970; Hendricks, 1966; Lumsden & Baird, 1985; Yoshida & Hendricks, 1970) . The reported increases in pressure amplitude at menses (Fig.4A) and in dysmenorrhea are consistent with the variations of JZ thickness (Fig.1B): the involvement of a thicker layer of smooth muscle in contractions at menses, or in dysmenorrhea, results in higher peak pressures.

The evolution of basal pressure, i.e., the baseline pressure reached between contractions, is less trivial. Investigators have consistently reported that baseline pressure is minimum at menses (close to atmospheric pressure, ~0-10 mmHg), and highest at ovulation (20-60 mmHg, Fig.4B) (Bulletti et al., 2000; Eskes et al., 1970; Hendricks, 1966; Yoshida & Hendricks, 1970). In a closed pressure vessel, the pressure P increases with wall tension σ (smooth muscle tone) and thickness t (JZ thickness), the mathematically simplest case being that of a thin-walled cylinder, where $P = \frac{\sigma t}{r}$. This relation predicts a high basal pressure at menses, and a low, relatively constant pressure during the remainder of the menstrual cycle. How do we resolve this conundrum?

We believe the key resides in how far the uterus can be considered a *closed* vessel. During menstruations, the cervix is slightly open and rigid; contractile waves move from the fundus towards the cervix; the uterine cavity equilibrates between contractions with the atmospheric pressure (Fig.4C,D): the organ is open towards the *outside*. In these conditions, the assumption of a closed vessel is not valid and smooth muscle tone does not induce a significant static pressure. On the contrary, during ovulation, the cervix is open but soft and the frequent contractile waves are directed from the cervix towards the fundus (Lyons et al., 1991) (Fig.4C). This constant upward peristaltic squeezing is very conspicuous in a recent high resolution 4D ultrasound video (Moliner et al., 2021). These peristaltic contractions likely act as intermittent sphincters that isolate the organ from atmospheric pressure: the ovulating uterus is an organ open towards the *inside*, i.e. towards the ovaries and the peritoneal cavity. In these conditions, myometrium tone can build up a static, basal pressure (Fig.4B,D).

For sperm to be propelled inside the uterine cavity at ovulation, cervico-fundal contractile waves have to overcome the basal pressure gradient between vagina and uterine cavity. We have previously shown that local peristaltic waves can induce luminal flow against a global pressure gradient (Amedzrovi Agbesi et al., 2024). The inherent motility of spermatozoa, and in particular rheotaxy (Miki & Clapham, 2013), does not appear necessary to overcome this pressure gradient: inert particles have been shown to move from the vagina to the uterine cavity during ovulation (Kunz et al., 1996). Dysregulation of wave direction during the menstrual cycle are a likely cause of infertility and of menses reflux, a primary mechanism of endometriosis development. We have recently shown on ex-vivo myometrial strips that wave direction can be controlled by ocytocin (Huon, Zig, et al., 2026), suggesting a key role of hormone receptor distribution inside the uterine muscle. Further research should aim at understanding the precise physiological mechanisms governing their directionality throughout the menstrual cycle.

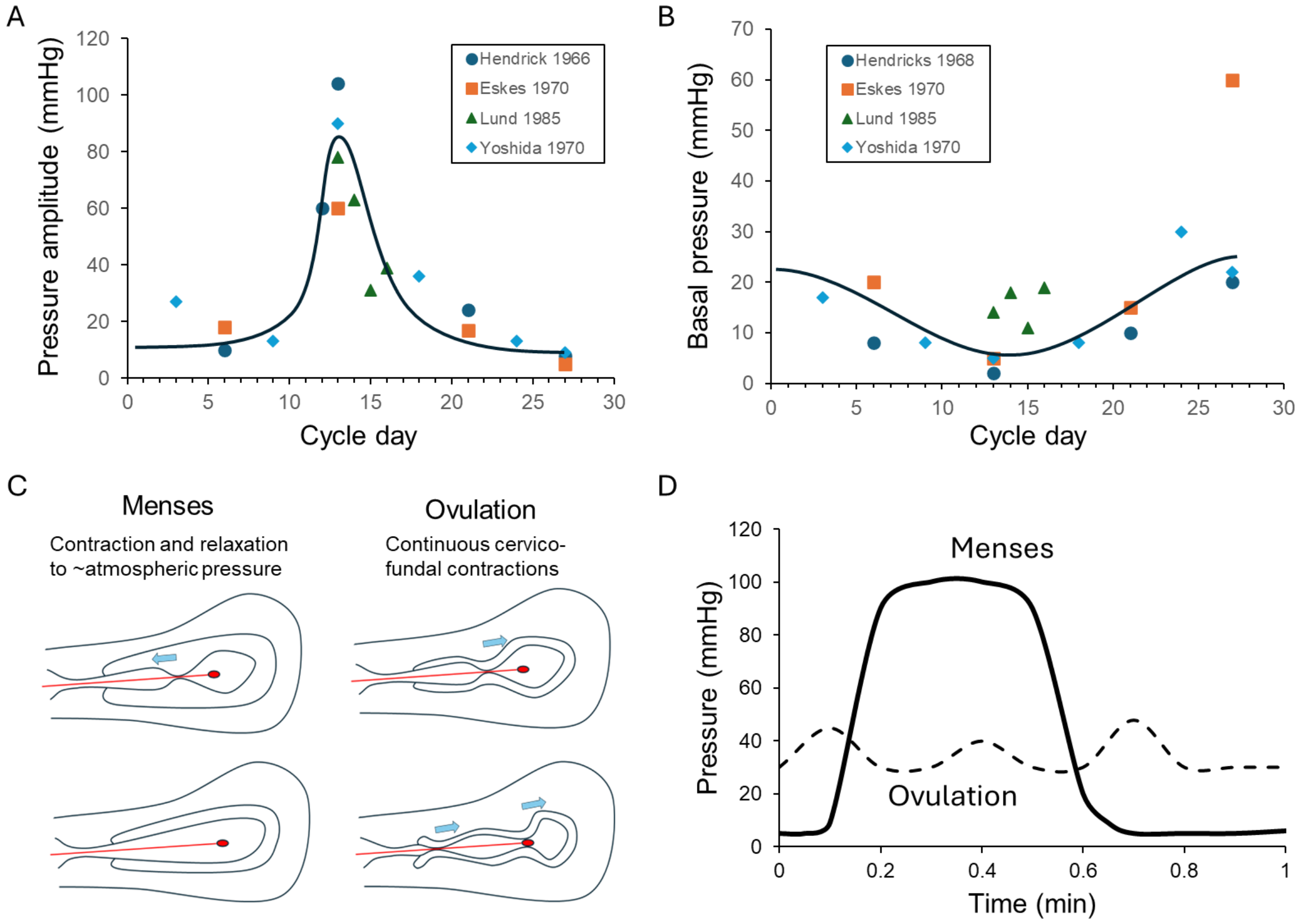

*Figure 4. Intra-uterine manometry through the menstrual cycle. A) Pressure amplitude is maximum at menses and minimal at ovulation. The evolution through the 3 days of menses has been reported by (Lumsden & Baird, 1985). B) Basal pressure follows at opposite trend: it is lowest at menses, and highest at ovulation. C) Schematic depiction of fundo-cervical infrequent uterine contractile waves at menses, and cervico-fundal frequent waves at ovulation; the latter can isolate the uterine cavity from atmospheric pressure by acting as intermittent sphincters. D) Resulting, schematic manometry profile at menses and ovulation.*

## Conclusion

We summarized existing knowledge on JZ statics and dynamics on different time scales ranging from a woman's life, to her menstrual cycle, and to individual contractions. Building on a previously formulated hypothesis, we emphasized that the JZ is not an anatomical structure, but a reflection of smooth muscle tone. The whole myometrium is potentially contractile, but its outer part is inhibited by sympathetic adrenergic innervation and its inner part is activated by a prostaglandin gradient from the endometrium and uterine lumen: a simple Hill-equation type model showed that a steep, step-like tone profile can emerge from this situation, reproducing the typical, well demarcated JZ. The involvement of a thicker smooth muscle layer in peristaltic contractions of the menstruating or diseased uterus gives rise to higher intra-uterine amplitude. We suggest that basal intra-uterine pressure measured by manometry is dependent on the direction and frequency of peristaltic waves: because these can act like sphincters, they determine in how far the organ is open to the outside (atmospheric pressure) or to the inside (Fallopian tubes, peritoneal cavity) during the menstrual cycle.

The myometrium is the thickest smooth muscle of the human body, but only a fraction of this thickness is necessary for its physiological function outside of pregnancy. This entails that most of the myometrium be “shut down”. Gynecological pathologies including severe dysmenorrhea, endometriosis and adenomyosis all appear intimately linked to a failure of the organism to ensure quiescence of most of the myometrium, resulting in “pregnancy-like” painful contractions outside of pregnancy.

## Acknowledgments

We gratefully acknowledge funding by the Fondation pour la Recherche sur l’Endométriose, project HYPERU, under the auspices of the Fondation pour la Recherche Médicale, project FRE202212016469. Emilie Huon was financed by an Ecole Doctorale de Physique de l’Ile-de-France (EDPIF) doctoral fellowship.